\documentstyle[aps,prc,twocolumn,floats,epsfig]{revtex}
\newcommand{\etal}{\emph{et al.}}
\begin{document}
\renewcommand{\thefootnote}{\arabic{footnote}}
\twocolumn[\columnwidth\textwidth\csname@twocolumnfalse\endcsname
\title{Shell Stabilization of Super- and Hyperheavy Nuclei 
       Without Magic Gaps}
\author{M. Bender,$^{1}$
        W. Nazarewicz,$^{2-4}$
        P.--G. Reinhard$^{5,6}$}
\address{$^1$Gesellschaft f\"ur Schwerionenforschung,
         Planckstrasse 1, D--64291 Darmstadt, Germany}
\address{$^2$Department of Physics and Astronomy,
         University of Tennessee, Knoxville, Tennessee 37996}
\address{$^3$Physics Division, Oak Ridge National Laboratory,
         P.O. Box 2008, Oak Ridge, Tennessee 37831}
\address{$^4$Institute of Theoretical Physics,
         Warsaw University,
         ul. Ho\.za 69, PL--00681, Warsaw, Poland}
\address{$^5$Institut f\"ur Theoretische Physik II,
         Universit\"at Erlangen-N\"urnberg,
         Staudtstrasse 7, D--91058 Erlangen, Germany}
\address{$^6$Joint Institute for Heavy Ion Research,
         Oak Ridge National Laboratory,
         P.\ O.\ Box 2008, Oak Ridge, Tennessee 37831}
\date{March 23, 2001}
\maketitle
%
%=======================================================================
%
\begin{abstract}
Quantum stabilization of superheavy elements is quantified in
terms of the shell-correction energy. We compute the shell correction
using self-consistent nuclear models: the non-relativistic
Skyrme-Hartree-Fock approach and the relativistic mean-field model,
for a number  of parametrizations. All the forces applied
predict a broad valley of shell stabilization around \mbox{$Z=120$} 
and \mbox{$N=172$}-184. 
We also predict two broad  regions of shell stabilization
in hyperheavy elements with \mbox{$N \approx 258$} and \mbox{$N \approx
308$}. Due to the large single-particle level density, shell corrections
in the superheavy elements differ markedly from those in lighter nuclei.
With increasing proton and neutron numbers, the regions of nuclei stabilized
by shell effects become poorly localized in particle number, and the
familiar pattern of shells separated by magic gaps is basically gone.
\end{abstract}
\addvspace{5mm}]
\narrowtext
%
%==========================================================================
%
The synthesis of superheavy elements (SHE) has been in the focus of
heavy-ion physics for  more than three decades. The last few years have
seen significant progress in our quest for reaching the region of 
long-lived SHE. Light isotopes of the elements \mbox{$Z=110$}-112 have
been safely established at GSI Darmstadt and JINR Dubna
\cite{Hof98a,Hof00a,Arm00b}. These isotopes are expected to be strongly
deformed thanks to the (predicted) deformed shells
\mbox{$Z=108$} and \mbox{$N=162$} (see Refs.~\cite{Cwi83,Mol94a,Smo95}
and references quoted therein). These new nuclides could be
unambiguously identified by their characteristic
$\alpha$-decay chains leading to
already known isotopes. Even heavier and more neutron-rich nuclides
have been announced just recently by GSI ($^{270}_{160}$110) \cite{Hof00b};
Dubna ($_{171}^{283}$112, $_{173-175}^{287-289}$114, and $_{176}^{292}$116)
\cite{Dubna}; and Berkeley ($_{175}^{293}$118) \cite{Berkeley}.
The $\alpha$-decay chains of those nuclei cannot be linked to any
known nuclides as they end with fissioning nuclei.
However, these results yet need to be confirmed \cite{Arm00b,Arm00a,Sch00}.

The mere existence of SHE relies on quantum mechanics. According to the 
classical liquid-drop picture, all superheavy nuclei should be
unstable against spontaneous fission --- due to the huge Coulomb repulsion.
However, additional stabilization of binding energy is possible
thanks to shell effects which generate local  minima
in the nuclear potential energy surface in the regions where the
level density around the Fermi level is lowered. The detailed
energy balance  between the local  minima is dictated by
the distribution of spherical single-particle orbitals.
In some cases the minima are sufficiently deep to stabilize the nucleus
against spontaneous fission;  the delay in the spontaneous fission
half-lives due to the shell effects can be as much as 15 orders of magnitude
for \mbox{$Z \gtrsim 106$} \cite{Mue88}.

The half-lives of the known isotopes of elements with \mbox{$Z>105$}
are predominantly limited by $\alpha$ decay and decrease from
0.9\,s for $^{263}_{157}$106 to 0.2\,ms for $^{277}_{165}{112}$.
These isotopes decay mostly by groups of successive $\alpha$ particles.
Although shell corrections strongly influence $Q_\alpha$ values, there
is no simple correlation between the magnitude of shell effects and
$\alpha$-decay half-lives \cite{Cwi96a}. For instance, if the shell
corrections are nearly constant in a broad region of particle numbers,
this will have very little influence on  $Q_\alpha$ values; hence
on $T_\alpha$.

The subject of the present paper is the shell stabilization quantified
in terms of the shell correction $E_{\rm shell}$. It is obtained
from decomposing the self-consistent binding energy $E_{\rm tot}$ as
\begin{equation}
\label{etot}
E_{\rm tot}
\approx  \tilde E + E_{\rm shell},
\end{equation}
where
$\tilde E$ is the average  energy that  changes smoothly with particle
number. (In microscopic-macroscopic approaches, $\tilde E$ is approximated
by the liquid drop or droplet model energy.)
The shell stabilization of SHE as predicted by macroscopic-microscopic
models has already  been  extensively discussed in the literature
\cite{Mol94a,Mol92a,Smo97a}. It is the aim of the present study to analyze  
shell corrections in the superheavy region
in the framework of self-consistent mean-field models.  To this end,
we apply the two most widely used approaches, namely the
Skyrme-Hartree-Fock (SHF) theory and the relativistic mean-field (RMF)
theory. (For a brief overview, see Ref.~\cite{commSHE}.)
In the previous paper by Kruppa {\em et al.}  \cite{Kru00a},
the shell energy extracted from self-consistent
single-particle spectra along a few selected isotopic and isotonic
chains was discussed. Here we present a large-scale survey of spherical
shell energies throughout the whole landscape of conceivable SHE.

There is a world of different parametrizations for SHF as well as RMF.
They agree more or less in their performance for stable nuclei
but can yield differing predictions in extrapolations.
For example, magic shell closures in SHE are at variance
\cite{Cwi96a,Kru00a,Rut97a,Ben99a}.
Interestingly, it has been concluded in Ref.~\cite{Kru00a}
that both the SHF and RMF calculations were
internally consistent. That is, all the Skyrme
models with conventional spin-orbit force predicted the strongest
spherical shell effect at \mbox{$N=184$} and \mbox{$Z=124$}, 126,
while all the RMF forces clearly preferred
\mbox{$N=172$} and \mbox{$Z=120$}, and SHF parameterizations
with relativistically extended spin-orbit interaction were in between.
%
%=========
%
\begin{figure*}[t!]
\centerline{\epsfig{file=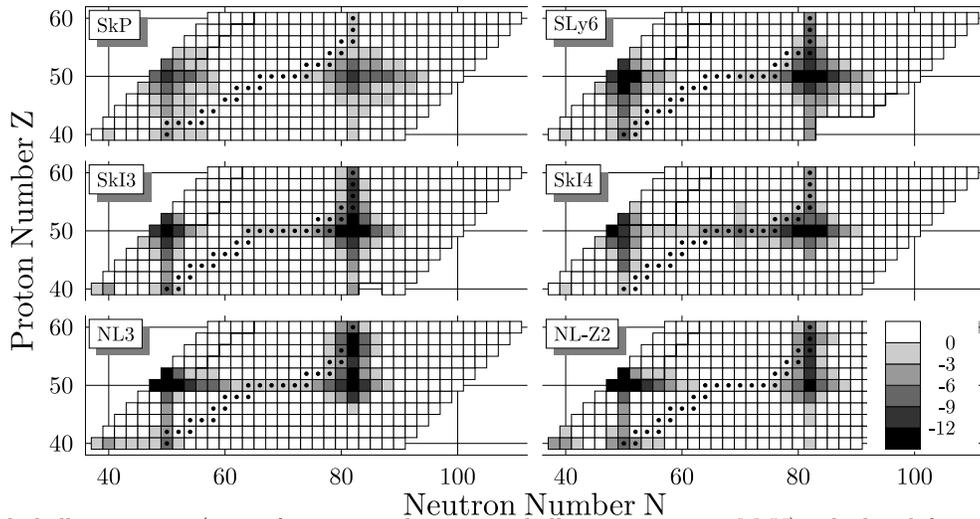}}
\caption{\label{fig:sn:sum} Total shell correction (sum of proton and neutron
shell corrections; in MeV) calculated for spherical even-even nuclei. The thick
solid lines denote two-particle drip lines. Black squares mark nuclei 
calculated to be stable with respect to $\beta$ decay. White color indicates
nuclei with positive shell corrections, black color denotes nuclei with 
$E_{\rm shell}$ beyond $-12$ MeV.
}
\end{figure*}
%
%=========
%

In view of these differences, we consider several parametrizations per
model. We use a selection of forces which have been found earlier to
represent the whole range of possible predictions for spherical shell
closures. For the RMF, we consider the parametrizations NL3 \cite{Lal97a}
and NL-Z2 \cite{Ben99a}. Both represent  recent fits which perform
very well with respect to global ground-state properties but differ
in detail. NL3 reproduces well isotopic trends, while NL-Z2 also fits the
electromagnetic nuclear form factor. For the SHF, we consider SkP
\cite{Dob84a} as a representative for a force with effective nucleon
mass \mbox{$m^*/m = 1$}, leading to a comparatively large density
of single-particle levels. All other SHF forces employed here
have smaller effective masses around \mbox{$m^*/m \approx 0.7$}.
SLy6 \cite{Cha98a} was adjusted with particular emphasis on isotopic
trends and neutron matter. SkI3 and SkI4 \cite{Rei95a} employ
an extended form of the spin-orbit force which was found to be
necessary for a description of isotope shifts
in heavy Pb isotopes. For SkI3, the extension was restricted to
map the spin-orbit structure of the RMF as closely as possible.

The shell energies are computed using the same prescription as
outlined in Ref.~\cite{Kru00a}.
This procedure, based on the Green's function approach to the level
density, is better suited for the calculation
of weakly bound systems than the traditional approach.
This is important in the context of nuclei considered here since
some of the predicted  regions of shell stability lie very close
to the proton drip line. In our calculations, we include a large space 
of single-particle states up to 50 MeV  above the Fermi energy.
Since most of these states are continuum (positive-energy) states,
the contribution from a particle gas (treated in the same numerical
box) has to be removed \cite{Kru00a}. Pairing correlations are ignored.

The calculations are restricted to spherical symmetry. Consequently, the 
calculated shell corrections represents in most cases an upper bound. 
In many cases, deformation does provide an additional binding, i.e.,
pulls to even stronger shell stabilization. Unfortunately, large-scale 
symmetry-unconstrained calculations of shell effects in the SHE are 
currently beyond our reach. This is because triaxial  \cite{Cwi96a} 
and reflection-asymmetric \cite{Ben98a} shapes must be considered
together with  more exotic topologies (e.g., bubble \cite{Dec99a},
toroidal, and rod structures) which might become favored for
the heaviest systems investigated here.

For nuclei up to \mbox{$Z=82$}, $E_{\rm shell}$ is always sharply peaked
at shell closures \cite{Rag84,Nil95}. Figure~\ref{fig:sn:sum} shows
an  example from the Sn region. The magic numbers \mbox{$Z=50$}, \mbox{$N=50$}, 
and \mbox{$N=82$} clearly stick out for all forces. One basically sees sharp
stripes along magic proton and neutron numbers. All forces show the same magic
numbers, but there are some differences in detail.
The overall strength of the shell effect seems to scale with effective
mass,  with SkP giving the smallest shell energies, while they are
most pronounced for NL3.
%
%=========
%
\begin{figure*}[ht!]
\centerline{\epsfig{file=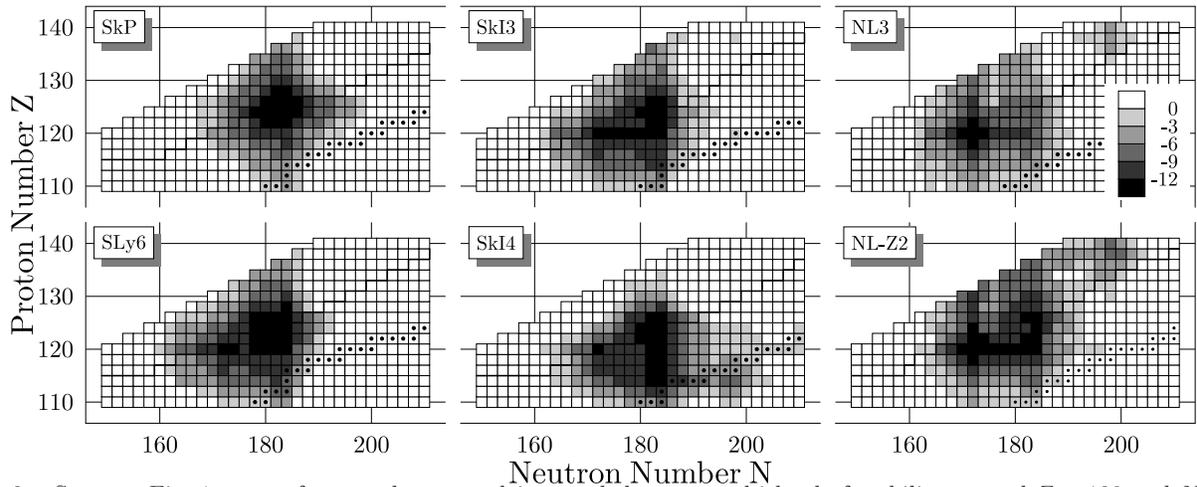}}
\caption{\label{fig:sum:all}
Same as Fig.~\protect\ref{fig:sn:sum} except for superheavy nuclei
around the expected island of stability around \mbox{$Z=120$} and
\mbox{$N=180$}. The scale for
$E_{\rm shell}$ is the same as in Fig.~\protect\ref{fig:sn:sum}.
See Ref.~\protect\cite{Kru00a} for the individual shell correction of protons
and neutrons along cuts through these maps.
}
\end{figure*}
%
%=========
%

The systematics of shell corrections change dramatically when
going to SHE; see Fig.~\ref{fig:sum:all}. Instead of narrow
stripes of large $E_{\rm shell}$ localized around magic numbers, all forces
employed predict a wide area of shell stabilization which spreads over all
shell closures predicted by the various forces. Unlike in normal nuclei,
large shell corrections in SHE  can appear slightly away from
shell closures. This was, in fact, already recognized in
macroscopic-microscopic models \cite{Mol94a,Mol92a,Smo97a}.
As a consequence, the significant differences seen in the prediction
of magic shells through various binding-energy indicators
(such as $\delta_{2q}$ \cite{Rut97a}) are much mellowed by the generally
softer pattern of the shell energy. We will discuss that aspect in more
detail below.
One of the common features seen in Fig.~\ref{fig:sum:all} is that the
region of nuclei with largest shell corrections
forms a triangle with the base at \mbox{$N=184$}
and the outer corner at \mbox{$^{292}_{172}120$}. This happens
because the existence of the \mbox{$N=172$} neutron subshell is strongly
coupled to the  proton subshell closure at \mbox{$Z=120$} \cite{Ben99a}.

At second glance, however, one also sees differences among various  
parameterizations. More stable nuclei are found above \mbox{$Z=120$} 
for SkP which predicts  the proton gap at \mbox{$Z=126$},
while the center of gravity is clearly shifted below \mbox{$Z=120$} for
SkI4 with its strong \mbox{$Z=114$} shell effect. The other forces reside
in between.  A somewhat different bias is also seen for the
extension in neutron direction. SkP predict strong
shell effects  for a number of nuclei with \mbox{$N > 184$},
while other forces fill basically the landscape between the two
magic numbers \mbox{$N=172$} and \mbox{$N=184$}.
There are also some differences concerning the overall area of the
stabilized region. For instance, NL3 makes it much smaller than all 
other forces. Of course, for a quantitative discussion, one needs  
to account for deformation effects which will serve to extend the island of
shell stabilization. For example, the well-known region of deformed
shell-stabilized SHE located around $^{270}_{162}$Hs$_{108}$
\cite{Cwi83,Mol94a,Smo95} is missing in Fig.~\ref{fig:sum:all},
as well as the deformed shell closure at
\mbox{$N=174$} \cite{Cwi99a,Ben00a}.
%
%=========
%
\begin{figure*}[ht!]
\centerline{\epsfig{file=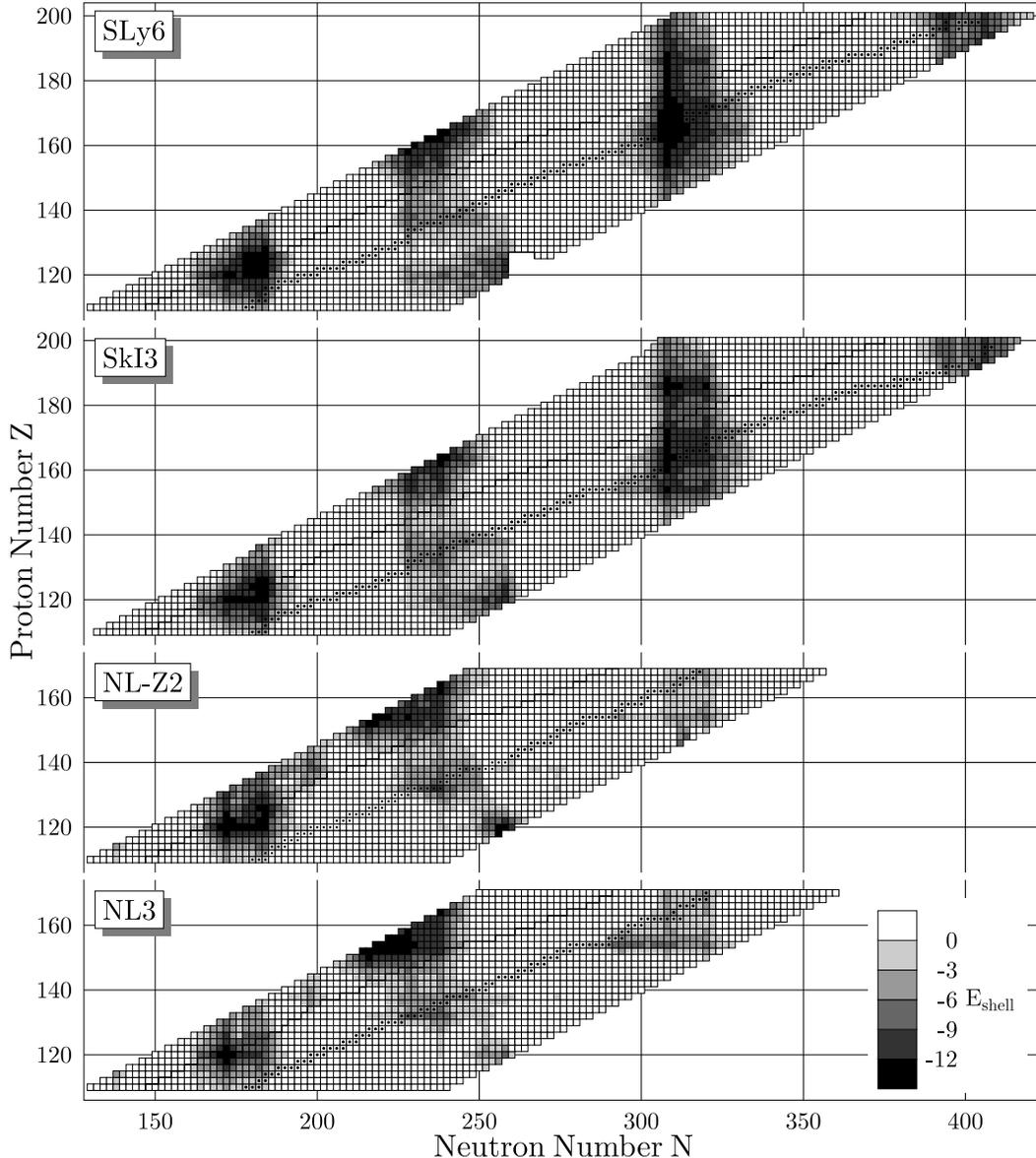}}
\caption{\label{fig:hyper:sum}
Same as in Figs.~\protect\ref{fig:sn:sum} and ~\protect\ref{fig:sum:all}
but for  larger neutron  and proton numbers
(up to \mbox{$Z=200$} for SHF and \mbox{$Z=170$} for RMF) calculated
with the subset of forces as indicated. The energy scale
is kept the same as in the other figures.
}
\end{figure*}
%
%=========
%

Figure \ref{fig:hyper:sum} takes a daring glance to even heavier
nuclei with \mbox{$Z>126$} (hyperheavy elements). There are several broad
valleys of spherical shell  stability showing up that extend
around the actual shell closures. This is a common feature of very
heavy nuclear systems. Differences between the forces, however,
grow dramatically in the hyperheavy region.
The upper limit for the plots is chosen to stay below the region
where the novel topologies, i.e.,  semi-bubbles
and bubbles, are believed to coexist \cite{Dec99a}.
  As the RMF predicts the lower border of
this transitional region at smaller values of $Z$ than SHF, we
set different upper limits of the displayed area in both approaches.
By inspecting Fig.~\ref{fig:hyper:sum} carefully, one can see
rather large differences in $E_{\rm shell}$ predicted in different
calculations. That is no surprise because fine details of shell
structure play an increasing role with increasing nuclear
size. Nevertheless, there still remains
an overall agreement concerning the position of the regions of stability,
around \mbox{$N=258$} and around \mbox{$N=308$}. The RMF
parametrizations are more pessimistically predicting only faint effects,
while  SHF produces a strong stabilization at \mbox{$N=308$}.

It is an open question whether for these hyperheavy elements the actual shell
corrections are sufficient to prevent (or significantly
slow down) spontaneous fission. One would expect
that, with increasing $Z$, the Coulomb force would act increasingly against
stability. In particular, the role of triaxial or reflection asymmetric degrees
of freedom must  be considered when assessing the stability of hyperheavy
nuclei to fission.
%
%=========
%
\begin{figure*}[ht!]
\centerline{\epsfig{file=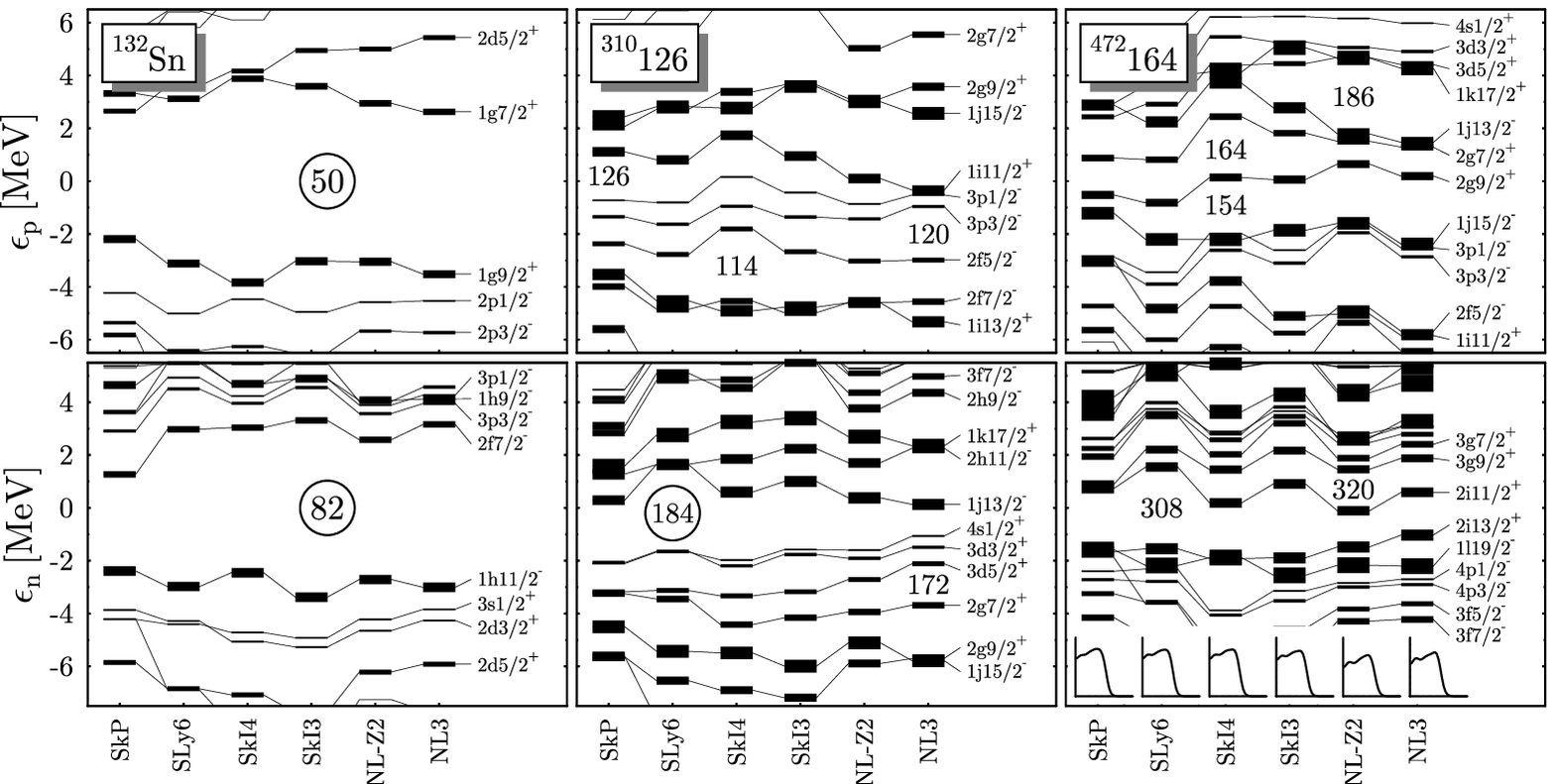}}
\caption{\label{fig:spectra}
Single-particle spectra of protons (top) and neutrons
(bottom) for $^{132}$Sn, $^{310}_{184}$126 and $^{472}_{308}$164.
The single-particle energies are taken relative to the Fermi energy  
predicted by SLy6. The line thickness of each level is proportional 
to the \mbox{$2j+1$} degeneracy of the state. The inset in 
the rightmost panel shows the predicted radial neutron distributions. 
No bubble structure is predicted  for $^{472}_{308}$164.
}
\end{figure*}
%
%=========
%

Figure~\ref{fig:spectra} shows the single-particle spectra for three
typical nuclei from the three regions of large shell correction
discussed in this paper. For $^{132}$Sn, proton and neutron magic gaps 
appear in all models. The patterns of single-particle levels are 
significantly different for the two regions of SHE.
Firstly, with increased mass, the overall level density grows as
$\propto A^{1/3}$. Secondly, no pronounced and
uniquely preferred energy gaps appear in the spectrum.
This shows that  shell closures which are to be associated with the
gaps in the spectrum are not robust in that region.
Tiny changes in, e.g., spin-orbit properties can shift the gaps
substantially; see \cite{Cwi96a,Ben99a} for a thorough discussion.
Protons are more sensitive in that respect than neutrons.

Interestingly, similar problems are encountered in atomic
calculations of the electron shell structure of SHE \cite{chemistry}.
Due to the large density of valence  electron shells, it is extremely difficult
to make robust predictions of chemical properties of SHE.

A close inspection of Fig.~\ref{fig:spectra} allows for a
rather good understanding of the shell-correction pattern discussed above.
For instance, in $^{310}_{184}$126  there appear low-$j$ single-particle
orbitals at the Fermi surface ($3p_{1/2}$ and $3p_{3/2}$ in the protons
between \mbox{$Z=120$} and 126 and $4s_{1/2}$, $3d_{3/2}$, and $3d_{5/2}$
in the neutrons between \mbox{$N=172$} and 184). The low (2$j$+1)
degeneracy of these shells gives rise to reduced single-particle
level densities; hence to a large negative shell-correction energy
for a whole range of neighboring nuclei.

For hyperheavy nuclei, the level density is even higher, and
no strong shell closures are predicted in most models.
It is only in SLy6 and SkI3 that a relative large \mbox{$N=308$} gap
is predicted, bounded by high-$j$ shells. This is consistent
with Fig.~\ref{fig:hyper:sum} which shows a rather strong neutron
shell effect at \mbox{$N=308$} for these two forces.

In summary, we have investigated the spherical  shell-stability 
of superheavy elements using state-of-the-art self-consistent models.
The investigation of the systematics of  shell energy
reveals a new feature when going to very heavy systems.  Beyond
\mbox{$Z=82$} and \mbox{$N=126$}, the familiar localization of the
shell effect at magic numbers is basically gone.
Instead, the theory predicts fairly wide areas of large shell
stabilization. Consequently, there is a good chance to
reach shell-stabilized SHE experimentally using a range of
beam-target combinations.

The disappearance of a familiar pattern of magic numbers and the appearance
of broad valleys of shell stability is due to (i) the rather large
single-particle level density, and (ii) the appearance of many low-$j$
shells around the Fermi level. This explains the
robustness of the shell correction in a rather large range of SHE and
at the same time the volatility of magic shell closures.

The results presented here have to be taken with a grain of salt.
Large shell correction is a neccesary, but not sufficient, condition
for the appearance of long-lived SHE. The calculations presented
in this paper serve as a starting point for subsequent
studies of deformed shell effects and fission barriers
in superheavy and hyperheavy nuclei.

%
%=======================================================================
%
\acknowledgements
This work was supported in part by
Bundesministerium f\"ur Bildung und Forschung (BMBF), Project
No.\ 06 ER 808; by Gesellschaft f\"ur Schwerionenforschung (GSI);
and by  the U.S. Department of Energy under Contract Nos.\ DE-FG02-96ER40963
(University of Tennessee), DE-FG05-87ER40361 (Joint Institute for Heavy
Ion Research), and DE-AC05-00OR22725 with UT-Battelle, LLC (Oak
Ridge National Laboratory).
%
%=======================================================================
%

\end{document}